\documentclass[aps,pre,twocolumn,groupedaddress]{revtex4-1}

\usepackage[pdftex]{graphicx}
\usepackage{bm}
\usepackage{amsmath}
\usepackage{amssymb}
\usepackage{amsfonts}
\usepackage{mathrsfs}

\begin{document}


\title{Coherence resonance for time-averaged measures}

\author{Go Uchida}
\affiliation{Department of Mechanical Systems Engineering, Tokyo Metropolitan University, Tokyo 1920397, Japan}


\date{\today}

\begin{abstract}
Noise can induce time order in the dynamics of nonlinear dynamical systems. 
For example, 
coherence resonance occurs in various neuron models driven by a noise. 
In studies of coherence resonance, 
ensemble-averaged measures of the coherence are often used. 
In the present study, 
we examine coherence resonance for time-averaged measures. 
For the examination, 
we use a Hodgkin-Huxley neuron model driven by a constant current and a noise. 
We firstly show that for large times, 
the neuron is in a stationary state irrespective of initial conditions of the neuron. 
We then show numerical evidence that 
in the stationary state, 
a given noise sample path uniquely determines the dynamics of the neuron. 
We then present numerical evidence suggesting that 
time-averaged coherence measures of the dynamics 
is independent of noise sample paths and 
is equal to ensemble-averaged coherence measures. 
On the basis of this property, 
we show that coherence resonance is not only a phenomenon related to ensemble-averaged measures  
but also a phenomenon that holds for time-averaged measures. 
\end{abstract}

\pacs{}

\maketitle

\section{Introduction}
Noise has unexpected effects on responses of nonlinear systems. 
Coherence resonance is an example. 
When an autonomous nonlinear system is driven by an external noise, 
regularity of its periodic response is maximal at a certain noise level.  
Coherence resonance was firstly found in a model of 
a simple autonomous system~\cite{Gang93}. 
Subsequently, 
it was found in 
the Plant model (a model for a bursting neuron)~\cite{Longtin97}, 
a FitzHugh-Nagumo model~\cite{Pikovsky97},  
a Hodgkin-Huxley model~\cite{Lee98},  
laser models~\cite{Dubbeldam99,Buldu01}, and 
a semiconductor superlattice model~\cite{Hizanidis08}. 
Coherence resonance was observed not only in models but also 
in experiments~\cite{Ushakov05,Arteaga07,Kabiraj15,Mompo18,Zhu19}.  

In studies of coherence resonance, 
measures of the regularity are often estimated by the ensemble average~\cite{Pikovsky97,Lee98,Luccioli06}. 
However, 
for comparison with experiments, 
it may be rather important that coherence resonance holds for time-averaged measures. 
Nevertheless, 
it is not necessarily clear whether coherence resonance also holds for time-averaged measures.  

In the present study, 
we examine coherence resonance for time-averaged measures. 
In the examination, 
we use a Hodgkin-Huxley neuron model driven by a constant current and a noise. 

This paper is organized as follows: 
In Sec.~\ref{sec:md}, 
we describe a Hodgkin-Huxley neuron model we use. 
The model is described by a set of stochastic differential equations. 
In Sec.~\ref{sec:pr}, 
some concepts of the theory of random dynamical systems are briefly explained. 
The theory of random dynamical systems 
provides a framework for pathwise analysis of stochastic differential equations. 
We also briefly review the dynamical behaviors of the Hodgkin-Huxley neuron. 
In Sec.~\ref{sec:brds}, 
we examine coherence resonance of the Hodgkin-Huxley neuron for time-averaged measures. 
In Sec.~\ref{sec:discuss}, 
we discuss the results.

\section{Model} \label{sec:md}
We use a Hodgkin-Huxley neuron model  
driven by a constant current and a noise. 
The electrophysiological activity of the Hodgkin-Huxley neuron is given by  
\begin{subequations}
\begin{eqnarray}
    C\frac{dv}{dt} &=& -g_{Na}m^{3} h (v-V_{Na}) - g_{K}n^{4} (v-V_{K}) 
                       \nonumber \\ & & -g_{L}(v-V_{L}) + I + \sigma \xi (t), \label{eqn:HH} \\
    \frac{dm}{dt} &=& \alpha_{m} (v) (1-m) - \beta_{m} (v) m, \label{eqn:m} \\
    \frac{dh}{dt} &=& \alpha_{h} (v) (1-h) - \beta_{h} (v) h, \label{eqn:h} \\
    \frac{dn}{dt} &=& \alpha_{n} (v) (1-n) - \beta_{n} (v) n, \label{eqn:n}
\end{eqnarray}
\end{subequations}
where $ v $ represents the membrane potential; 
$ C $ is the membrane capacitance; 
$ g_{Na} $, $ g_{K} $, and $ g_{L} $ are the maximum conductance for 
sodium ion, potassium ion, and leakage channels, respectively; 
$ V_{Na} $, $ V_{K} $, and $ V_{L} $ are the reversal potentials; 
$ m $, $ n $, and $ h $ are the gating variables. 
In Eqs.~(\ref{eqn:m}) to (\ref{eqn:n}), 
$ \alpha_{m}(v) $, $ \alpha_{h}(v) $, $ \alpha_{n}(v) $, 
$ \beta_{m}(v) $, $ \beta_{h}(v) $ and $ \beta_{n}(v) $ are 
the voltage-dependent rate constants. 
The voltage-dependent rate constants have the form~\cite{Abbott90}: 
\begin{subequations}
\begin{eqnarray}
    \alpha_{m}(v) &=& \frac{0.1(v+40)}{1-\exp[-(v+40)/10 ]}, \label{eqn:amHHsqd}\\ 
    \beta_{m}(v) &=& 4\exp[-(v+65)/18 ], \\
    \alpha_{h}(v) &=& 0.07\exp[-(v+65)/20], \\
    \beta_{h}(v) &=& \frac{1}{1+\exp[-(v+35)/10]}, \\
    \alpha_{n}(v) &=& \frac{0.01(v+55)}{1-\exp[-(v+55)/10]}, \\
    \beta_{n}(v) &=& 0.125\exp[-(v+65)/80].  \label{eqn:bnHHsqd}
\end{eqnarray}
\end{subequations}
In Eq.~(\ref{eqn:HH}), 
$ I $ represents a constant external current and 
$ \xi (t) $ represents a Gaussian white noise: $ \left\langle \xi (t) \right\rangle = 0 $ and 
$ \left\langle \xi (t) \xi (t') \right\rangle = \delta (t-t') $. 
The symbol $ \sigma $ represents the amplitude of the noise.

\section{Preliminaries} \label{sec:pr}
\subsection{Some concepts of the theory of random dynamical systems}    \label{app:erds}
In the present study, 
we use some concepts of the theory of random dynamical systems.  
The theory of random dynamical systems 
provides a framework for pathwise analysis of stochastic differential equations.  
Here, 
a brief explanation of the concepts is given. 
More detailed and rigorous descriptions of the concepts can be found in~\cite{Arnold98}.

\subsubsection{Pullback method, random attractor, and invariant measure}
Here, 
we focus on the systems described by stochastic differential equations as random dynamical systems, 
although random dynamical systems include systems not described by stochastic differential equations. 

We assume that the dynamics of a system is described by a stochastic differential equation:  
\begin{equation}
    \frac{dx}{dt} = f(x) + \sigma \xi (t).     \label{eq:irexamg}
\end{equation}
We denote a formal solution of Eq.~(\ref{eq:irexamg}) for a given noise sample path $ \omega $ as $ x(t,\omega ) $ and 
the initial condition as $ x_{0} $. 
In the field of random dynamical systems, 
the solution $ x(t,\omega ) $ is expressed as $ \varphi (t,\omega ) x_{0} $ using a map $ \varphi (t,\omega ) $.  

A random attractor $ \mathscr{A} $ is defined as $ \mathscr{A} = \left\{ A(\omega ) \right\} _{\omega \in \Omega } $ 
where $ \Omega $ represents the set of all $ \omega $, 
and $ A(\omega ) $ is defined as a $ \varphi $-invariant set that attracts, 
in a pullback sense, 
all points in a region of the phase space.  
Here, 
$ \varphi $-invariant means that the following equation holds: 
\begin{equation}
    \varphi (t, \omega ) A(\omega ) = A(\theta _{t} \omega ),    \label{eq:defatr}
\end{equation} 
where $ \theta _{t} $ represents the shift operator and 
maps $ \xi (s;\omega ) $ to $ \xi (s+t;\omega ) $. 
It is known that for a Gaussian white noise, 
$ \theta _{t} $ is a bijection from $ \Omega $ to $ \Omega $. 

The pullback means $ \lim_{t \to \infty } \varphi (t,\theta _{-t} \omega ) x_{0} $. 
This corresponds to characterizing the asymptotic behavior of the system by 
the time evolution from $ t = -\infty $ to $ t = 0 $ instead of 
the time evolution from $ t = 0 $ to $ t = \infty $. 
The reason for using the pullback will become clear in the next section (Sec.~\ref{sssec:example}). 

Random dynamical systems do not necessarily have random attractors. 
A random dynamical system possesses a random attractor 
if 
all trajectories starting at $ t' = -\infty $ are 
within a bounded region $ B(\omega ) $ at the time $ t = 0 $~\cite{Arnold98,Crauel99,Caraballo17}.  

From the point of view of probability theory, 
if a random dynamical system has a random attractor, 
then the system almost surely has 
a $ \varphi $-invariant conditional probability given a noise sample path $ \omega $~\cite{Arnold98,Crauel99}:  
\begin{equation}
    \mu _{\omega } \left(A(\omega ) \right) = 1,    \label{eq:rasuppA}
\end{equation}
where $ \mu _{\omega } $ represents the $ \varphi $-invariant conditional probability 
given a noise sample path $ \omega $. 
The probability $ \mu _{\omega } $ is called an invariant measure. 

For a random dynamical system,   
the invariant measure and the stationary solution of the Fokker-Planck equation for the system have 
one-to-one correspondence~\cite{Arnold98,Crauel99}: 
\begin{eqnarray}
    \mathbb{E}_{\omega } \left[ \mu _{\omega } (dx) \right] &=& \rho (x)dx,   \label{eq:murho} \\
    \lim_{t \to \infty } \varphi (t,\theta _{-t} \omega ) \rho (x)dx &=& \mu _{\omega }(dx), 
\end{eqnarray}
where $ \mathbb{E}_{\omega } $ represents the expectation and 
$ \rho (x) $ is a solution of the equation:  
\begin{equation}
    \left[ -\frac{\partial }{\partial x} f(x) + \frac{{\sigma }^2}{2}\frac{{\partial }^2}{\partial x^2} \right] \rho (x) = 0.  
\end{equation}

The point in this section is that 
for large times, 
the system is in a stationary state irrespective of initial conditions if 
the system has a random attractor.

\subsubsection{Example}    \label{sssec:example}
Here, 
a simple example is provided 
to facilitate understanding of the concepts explained in the previous section. 
The theory of random dynamical systems is applicable not only to 
nonlinear systems with a noise but also to linear systems with a noise. 
For simplicity, 
we use a linear system with a noise as an example.  
The example we use is as follows:  
\begin{equation}
    \frac{dx}{dt} = -\gamma x + \sigma \xi (t),    \label{eq:irexam}
\end{equation}
where $ \gamma $ is a positive constant. 

For a given noise sample path $ \omega $, 
a formal solution of Eq.~(\ref{eq:irexam}) is given by
\begin{equation}
    \varphi (t,\omega ) x_{0} = e^{-\gamma t} x_{0} + \sigma \int _{0}^{t} e^{-\gamma (t-s)} \xi (s;\omega ) ds.    \label{eq:irexamsol}
\end{equation}
From Eq.~(\ref{eq:irexamsol}), 
we can see that $ \lim_{t \to \infty } \varphi (t,\omega ) x_{0} $ is indeterminate. 
This makes it difficult to characterize asymptotic behaviors of the system. 
However, 
this difficulty is overcome by using the pullback. 
For the system given by Eq.~(\ref{eq:irexam}), 
we have the pullback:  
\begin{equation}
    \lim_{t \to \infty } \varphi (t,\theta _{-t} \omega ) x_{0} = \sigma \int_{-\infty }^{0} e^{\gamma s} \xi (s;\omega ) ds.    \label{eq:aat} 
\end{equation}
This pullback is bounded because $ \varphi (t,\omega ) x_{0} $ is an Ornstein-Uhlenbeck process. 
We denote the limit in Eq.~(\ref{eq:aat}) as $ x^{*}(\omega ) $. 

The system given by Eq.~(\ref{eq:irexam}) has a random attractor because 
the pullback is bounded. 
In addition, 
all solutions converge to $ x^{*}(\omega ) $ irrespective of initial conditions and  
$ x^{*}(\omega ) $ is $ \varphi $-invariant (see Appendix~\ref{app:inv}): 
\begin{equation}
    \varphi (t, \omega ) x^{*}(\omega ) = x^{*}(\theta _{t} \omega ).    \label{eq:auinv}
\end{equation}
The random attractor 
$ A(\omega ) $ is given by 
\begin{equation}
    A(\omega ) = \{ x^{*}(\omega ) \}.    \label{eq:raexam} 
\end{equation} 
If a random attractor is the family of singletons, 
the attractor is called a random point attractor. 

From Eqs.~(\ref{eq:rasuppA}) and (\ref{eq:raexam}), 
an invariant measure for the system given by Eq.~(\ref{eq:irexam}) is given by 
\begin{equation}
    \mu _{\omega } = \delta _{x^{*}(\omega )}.    \label{eqn:ltlcp}  
\end{equation}  
The measure $ \delta _{x^{*}(\omega )} $ is called a random Dirac measure and is given by 
$ \delta _{x^{*}(\omega )} = \delta \left( x-x^{*}(\omega ) \right) dx $. 

The stationary solution of the Fokker-Planck equation for the system described by Eq.~(\ref{eq:irexam}) 
is given by 
\begin{equation}
    \rho (x) = \frac{1}{\sqrt{\pi } \sigma } \exp \left( -\frac{x^{2}}{\sigma ^{2}} \right). \label{eq:rhoexp}
\end{equation}
Thus, 
from Eq.~(\ref{eq:murho}), 
we have 
\begin{equation}
    \mathbb{E}_{\omega } \left[ \delta _{x^{*}(\omega )} \right] = \frac{1}{\sqrt{\pi } \sigma } \exp \left( -\frac{x^{2}}{\sigma ^{2}} \right) dx. 
\end{equation}

The point in this section is that if 
a random attractor of a system is a random point attractor, 
in the stationary state, 
a given noise sample path uniquely determines the dynamics of the system.

\subsection{Dynamical behaviors of the Hodgkin-Huxley neuron}    \label{sec:DB}
Here, 
we briefly review the dynamics of the Hodgkin-Huxley neuron given by Eqs.~(\ref{eqn:HH}) to (\ref{eqn:bnHHsqd}) 
to clarify the regions of $ I $ and $ \sigma $ where coherence resonance occurs. 

When $ \sigma = 0 $, 
in the Hodgkin-Huxley neuron given by 
Eqs.~(\ref{eqn:HH}) to (\ref{eqn:bnHHsqd}), 
a saddle-node bifurcation of periodic orbits occurs at 
$ I = 6.23 $ $ \mu $A$/$cm$^{2}$ and 
a Hopf bifurcation occurs at 
$ I = 9.78 $ $ \mu $A$/$cm$^{2}$. 
For $ I < 6.23 $ $ \mu $A$/$cm$^{2}$, 
a stable fixed point is the only attractor. 
For $ 6.23 $ $ \mu $A$/$cm$^{2}$ $< I < 9.78 $  $ \mu $A$/$cm$^{2}$, 
a stable fixed point, a stable limit cycle, and an unstable limit cycle 
coexist and the unstable limit cycle is the separatrix between 
the stable fixed point and the stable limit cycle. 
For $ 9.78 $ $ \mu $A$/$cm$^{2}$ $< I $, 
a stable limit cycle is the only attractor. 

When $ I < 6.23 $ $ \mu $A$/$cm$^{2}$, 
moderate to high amplitude noise induces stochastic firing in the Hodgkin-Huxley neuron given by 
Eqs.~(\ref{eqn:HH}) to (\ref{eqn:bnHHsqd}). 
This stochastic firing is based on excitable dynamics: 
noise makes the neuron an excursion into 
the region of the limit cycle. 
In this region of the input parameters, 
coherence resonance is observed~\cite{Luccioli06}.

\section{Results} \label{sec:brds} 
In this section, 
we firstly show analytically that the Hodgkin-Huxley neuron has a random attractor and 
then show numerically that 
the attractor is a random point attractor. 
We then characterize the dynamics in the stationary state  
and examine coherence resonance for time-averaged measures.

\subsection{Existence of a random attractor}    \label{ss:EAI}
The Hodgkin-Huxley neuron given by Eqs.~(\ref{eqn:HH}) to (\ref{eqn:bnHHsqd}) has a random attractor and thus 
for large times, 
the neuron is in a stationary state irrespective of initial conditions.  
The gating variables $ m $, $ h $,
and $ n $ are always bounded between zero and one irrespective of the noise sample path $ \omega $. 
In addition, 
the pullback of $ v $ is also bounded (see Appendix~\ref{app:bv}). 
Here, 
it is worth noting that 
the existence of a random attractor does not depend on the values of $ I $ and $ \sigma $.

\subsection{Structure of the random attractor}
We next show numerical evidence that the random attractor is a random point attractor  
in a parameter range where coherence resonance occurs. 
The initial conditions for the numerical calculations are given in a grid form over a wide region of the phase space: 
the initial conditions are all possible combinations of 
$v_{i} = V_{K} + i(V_{Na} - V_{K})/5 \, (i=0,1,\cdots ,5) $, 
$ m_{j} = 0 + j/4 \, (j=0,1,\cdots, 4) $, 
$ h_{k} = 0 + k/4 \, (k=0,1,\cdots, 4) $,
and $ n_{l} = 0 + l/4 \, (l=0,1,\cdots, 4) $; 
the number of the initial conditions are 750. 
The values of the model parameters we use 
in calculations are shown in Table~\ref{tab:paramvals}. 
\begin{table}
\caption{\label{tab:paramvals} Values of the model parameters}
\begin{ruledtabular}
\begin{tabular}{ll}
Parameters & Values \\ \hline
$ C $ & 1 $\mu $F$/$cm$^{2}$ \\ 
$ g_{Na} $ & 120 mS$/$cm$^{2}$ \\ 
$ g_{K} $ & 36 mS$/$cm$^{2}$ \\ 
$ g_{L} $ & 0.3 mS$/$cm$^{2}$ \\
$ V_{Na} $ & 50 mV \\
$ V_{K} $ & -77 mV \\
$ V_{L} $ & -54.4 mV 
\end{tabular}
\end{ruledtabular}
\end{table}

\begin{figure}
\includegraphics[width=8.6cm]{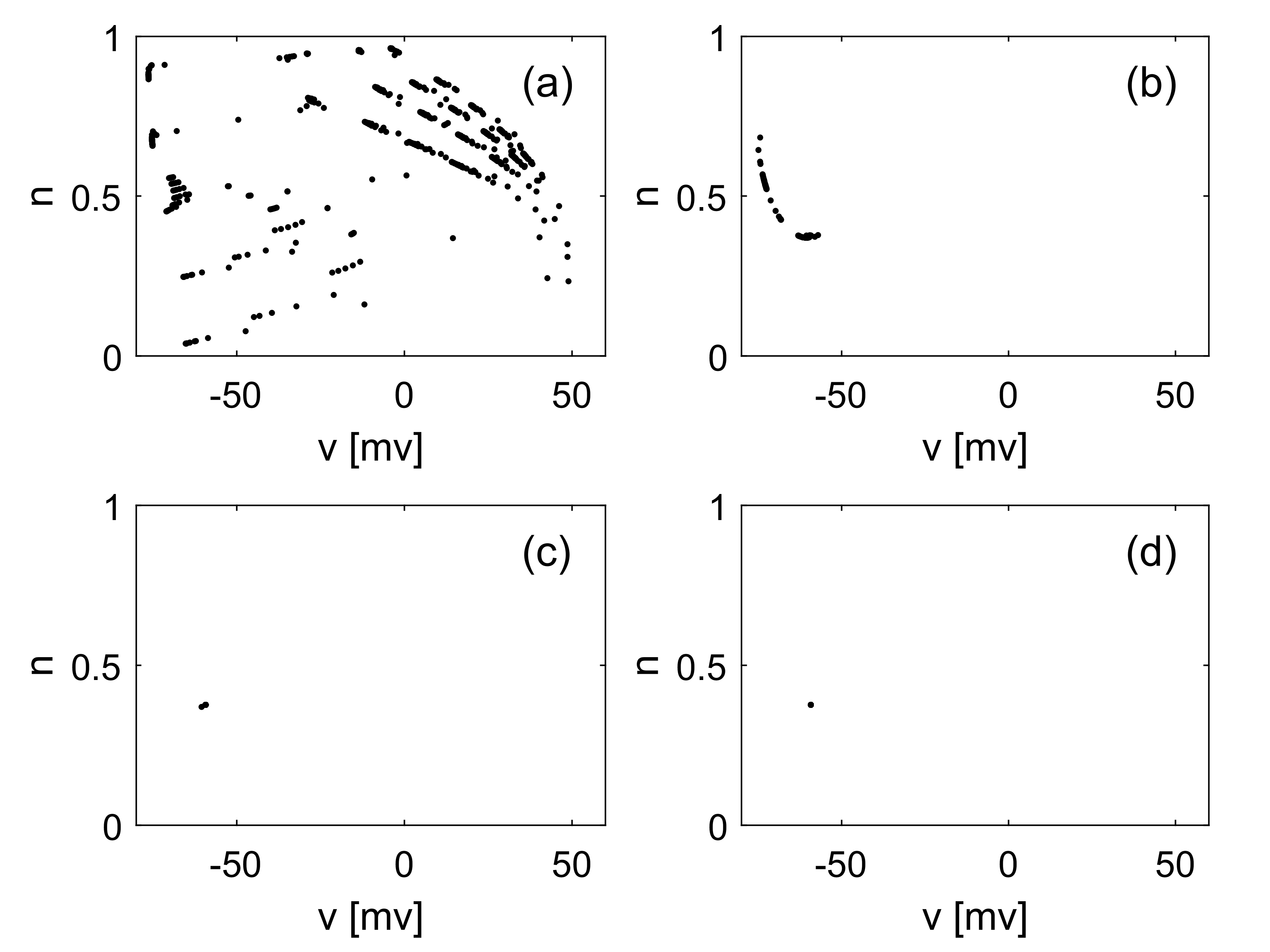}%
\caption{\label{fig:ssimn6p3nv} Pullbacks of the Hodgkin-Huxley neuron for a sample path of $ \xi (t) $. 
$ I = 6.2 $ $ \mu $A$/$cm$^{2}$ and $ \sigma = 10 $ $ \mu $A$ \cdot $ ms$^{1/2}/$cm$^{2}$. 
(a) $ t = 1 $ ms. (b) $ t = 200 $ ms. (c) $ t = 500 $ ms. (d) $ t = 2\, 000 $ ms. 
}
\end{figure}
In the present study, 
we fix the value of $ I $ to $ 6.2 $ $ \mu $A$/$cm$^{2}$ and 
change the value of $ \sigma $. 
We firstly set 
$ \sigma = 10 $  $ \mu $A$ \cdot $ ms$^{1/2}/$cm$^{2}$.  
Figure~\ref{fig:ssimn6p3nv} shows the pullbacks 
projected onto $ v $-$ n $ plane.  
The projected points converge to 
a single point irrespective of the initial conditions. 
The result is the same for the pullbacks projected onto the other planes. 
These results mean that the trajectories in the phase space converge to 
a single trajectory irrespective of the initial conditions. 
For other sample paths of $ \xi (t) $,  
we have the same result. 
These results suggest that 
the random attractor is a random point attractor and thus 
for a given noise sample path, 
the dynamics of the neuron is uniquely determined in the stationary state. 

For other values ($ 6 $, $ 8 $, $ 20 $, $ 40 $, $ 60 $, $ 80 $, and $ 100 $ $ \mu $A$ \cdot $ ms$^{1/2}/$cm$^{2}$) 
of $ \sigma $, 
we obtain the same result: the trajectories in the phase space converge to 
a single trajectory irrespective of the initial conditions. 
This result suggests that the random attractor is also a random point attractor for 
those values of $ \sigma $ and thus 
for a given noise sample path, 
the dynamics of the neuron is uniquely determined in the stationary state.

\subsection{Dynamics in the stationary state}
Unfortunately, 
the theory of random dynamical systems does not provide 
analytical methods to obtain further insight into the dynamics 
in the stationary state. 

\begin{figure}
\includegraphics[width=8.6cm]{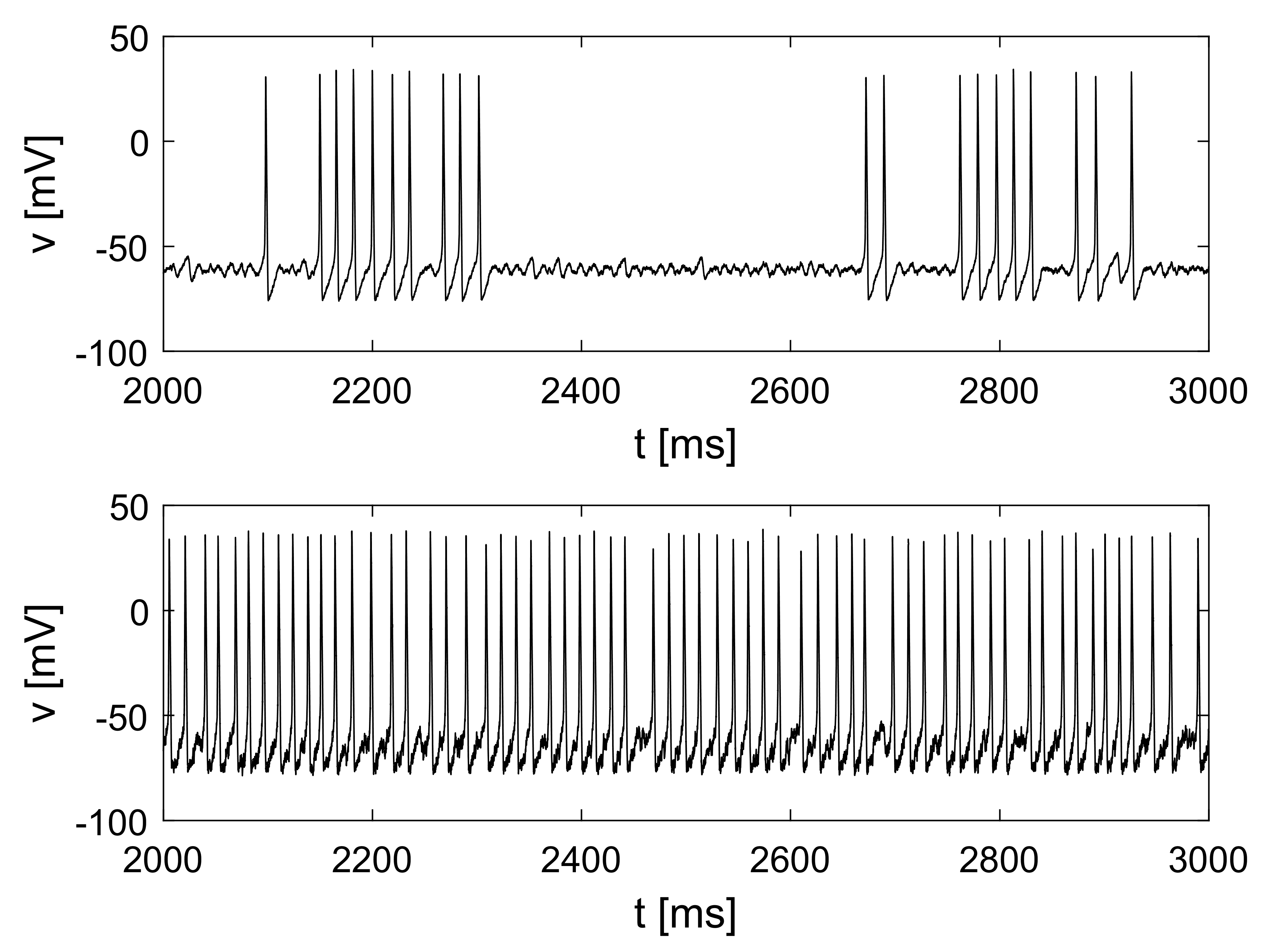}%
\caption{\label{fig:timn6p3} Time courses of the membrane potentials. 
$ I = 6.2 $ $ \mu $A$/$cm$^{2}$. Top panel: $ \sigma = 10 $ $ \mu $A$ \cdot $ ms$^{1/2}/$cm$^{2}$, 
Bottom panel: $ \sigma = 40 $ $ \mu $A$ \cdot $ ms$^{1/2}/$cm$^{2}$. 
}
\end{figure}
Figure~\ref{fig:timn6p3} shows the time courses of $ v $ 
in the stationary state 
for 
$ \sigma = 10 $  $ \mu $A$ \cdot $ ms$^{1/2}/$cm$^{2}$ and $ \sigma = 40 $  $ \mu $A$ \cdot $ ms$^{1/2}/$cm$^{2}$. 
We can see that the membrane potentials show 
intermittent oscillations for both values of $ \sigma $. 
Interestingly, 
however, 
the oscillations appear to be more regular at $ \sigma = 40 $  $ \mu $A$ \cdot $ ms$^{1/2}/$cm$^{2}$ than 
at $ \sigma = 10 $  $ \mu $A$ \cdot $ ms$^{1/2}/$cm$^{2}$. 

We then quantify the irregularity of the intermittent oscillation.
We cut the random point attractor by a section $ v = -40 $ mV, $ 0.1 \le m \le 0.4 $, $ 0.2 \le h \le 0.8 $, $ 0.1 \le n \le 0.6 $ (Poincar{\'e} section).   
For a given noise sample path $ \omega $, 
we denote the $ n $-th recurrence time of the neuron to the Poincar{\'e} section as $ T_{n} (\omega ) $. 
We define the time-averaged irregularity of an oscillation as  
\begin{equation}
    \overline{R}(\omega ) = \frac{\sqrt{\overline{T^{2}} (\omega ) - \left[ \overline{T} (\omega ) \right] ^{2}}}{\overline{T} (\omega )},    \label{eq:CV}  
\end{equation}
where $ \overline{R}(\omega ) $ represents the time-averaged irregularity of an oscillation. 
In Eq.~(\ref{eq:CV}), $ \overline{T} (\omega ) $ and $ \overline{T^{2}} (\omega ) $ are given by 
\begin{eqnarray}
    \overline{T} (\omega ) &=& \lim_{N \to \infty } \frac{1}{N} \sum_{n=1}^{N} T_{n} (\omega ),    \\
    \overline{T^{2}} (\omega ) &=& \lim_{N \to \infty } \frac{1}{N} \sum_{n=1}^{N} T_{n}^{2} (\omega ). 
\end{eqnarray}
If an attractor is a deterministic limit cycle, 
the recurrence time is a constant regardless of $ n $ 
and thus  
$ \overline{R}(\omega ) = 0 $.  
On the other hand, 
when the recurrence time follows an exponential distribution,  
$ \overline{R}(\omega ) = 1 $. 

\begin{figure}
\includegraphics[width=8.6cm]{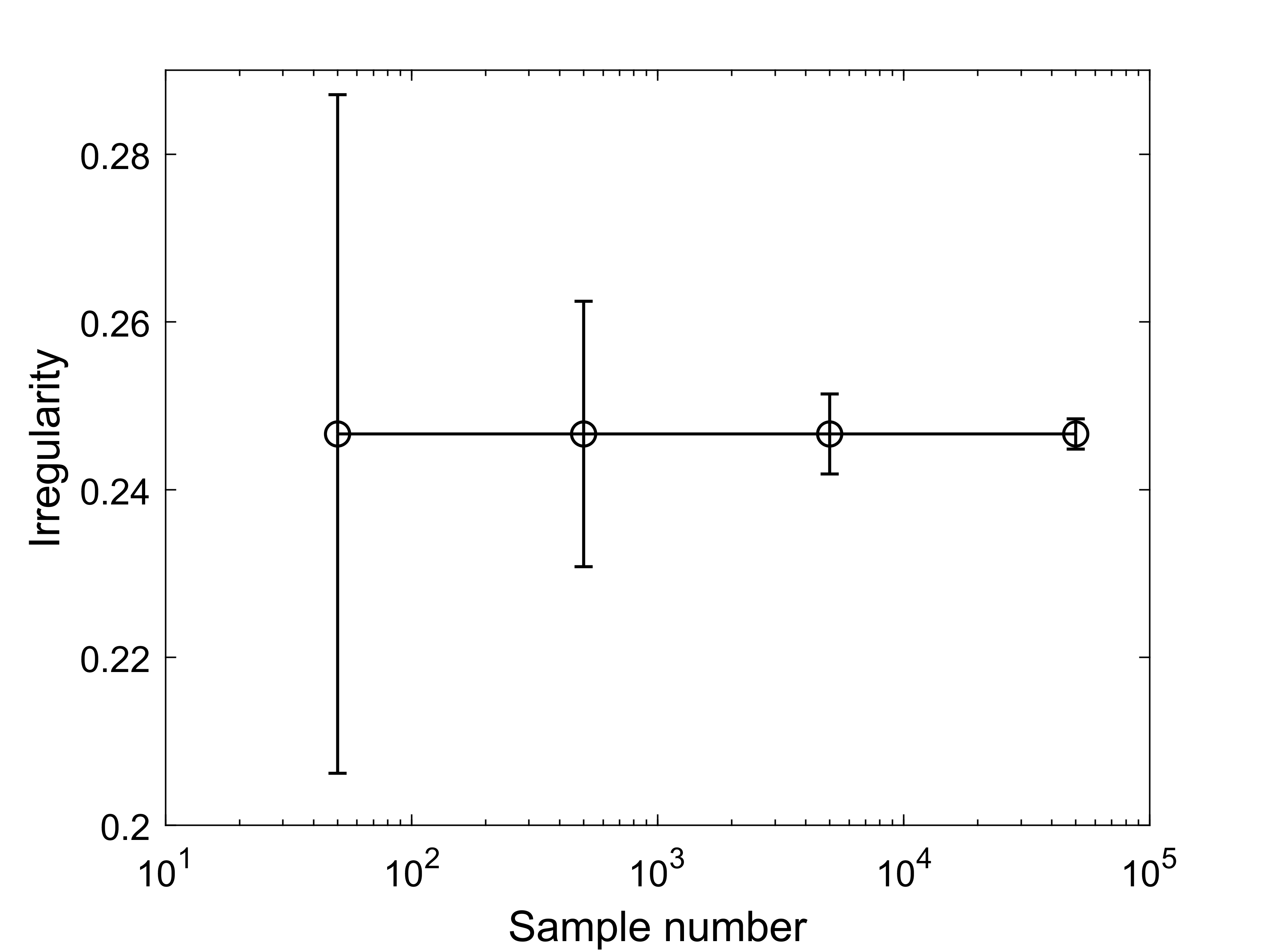}%
\caption{\label{fig:dCoS6p3te} The dependence of the difference between the time-averaged and the ensemble-averaged irregularity 
on the sample number of the recurrence times. 
$ I = 6.2 $ $ \mu $A$/$cm$^{2}$ and $ \sigma = 40 $  $ \mu $A$ \cdot $ ms$^{1/2}/$cm$^{2}$. 
The open circles represent the ensemble-averaged irregularity.
The error bars represent standard deviation $ S_{R} (N) $.  
The solid line is a guide for eyes.}
\end{figure}
Here, 
we also define the ensemble-averaged irregularity of an oscillation $ \left\langle R(\omega ) \right\rangle $: 
\begin{equation}
    \left\langle R(\omega ) \right\rangle = \frac{\sqrt{\left\langle T_{1}^{2} (\omega ) \right\rangle- \left\langle T_{1} (\omega ) \right\rangle^{2}}}{\left\langle T_{1} (\omega ) \right\rangle}.    \label{eq:eaCV}  
\end{equation} 
Figure~\ref{fig:dCoS6p3te} shows the sample number $ N $ dependence of the standard deviation $ S_{R} (N) $. 
Here, 
the standard deviation $ S_{R} (N) $ is given by 
\begin{equation}
    S_{R} (N) = \sqrt{ \left\langle \left( \widehat{\overline{R}}(N, \omega ) - \left\langle R(\omega ) \right\rangle \right) ^{2} \right\rangle },  
\end{equation}
where $ \widehat{\overline{R}}(N, \omega ) $ is an estimate of $ \overline{R}(\omega ) $ and is given by 
\begin{equation}
    \widehat{\overline{R}}(N, \omega ) = \frac{\sqrt{\widehat{\overline{T^{2}}} (N, \omega ) - \left[ \widehat{\overline{T}} (N, \omega ) \right] ^{2}}}{\widehat{\overline{T}} (N, \omega )} .
\end{equation}
In this equation, 
$ \widehat{\overline{T}} (N, \omega ) $ and $ \widehat{\overline{T^{2}}} (N, \omega ) $ are given by 
\begin{eqnarray}
    \widehat{\overline{T}} (N, \omega ) &=& \frac{1}{N} \sum_{n=1}^{N} T_{n} (\omega ),    \\
    \widehat{\overline{T^{2}}} (N, \omega ) &=& \frac{1}{N} \sum_{n=1}^{N} T_{n}^{2} (\omega ). 
\end{eqnarray}
From Fig.~\ref{fig:dCoS6p3te}, 
we can see that  
the standard deviation monotonically decreases as the sample number increases. 
This result suggests that $ \overline{R}(\omega ) $ is independent of $ \omega $ and 
is equal to $ \left\langle R(\omega ) \right\rangle $. 

In the following, 
we simply denote the irregularity as $ R $, 
but the estimation is based on the time average. 
As we have just shown, 
for $ \sigma = 40 $  $ \mu $A$ \cdot $ ms$^{1/2}/$cm$^{2}$, 
$ R = 0.2465 $. 
On the other hand, 
for $ \sigma = 10 $  $ \mu $A$ \cdot $ ms$^{1/2}/$cm$^{2}$, 
$ R = 1.1385 $. 
The value of $ R $ is smaller for $ \sigma = 40 $  $ \mu $A$ \cdot $ ms$^{1/2}/$cm$^{2}$ 
than for $ \sigma = 10 $  $ \mu $A$ \cdot $ ms$^{1/2}/$cm$^{2}$. 
This result is consistent with the result obtained by the comparison in Figure~\ref{fig:timn6p3}. 

\begin{figure}
\includegraphics[width=8.6cm]{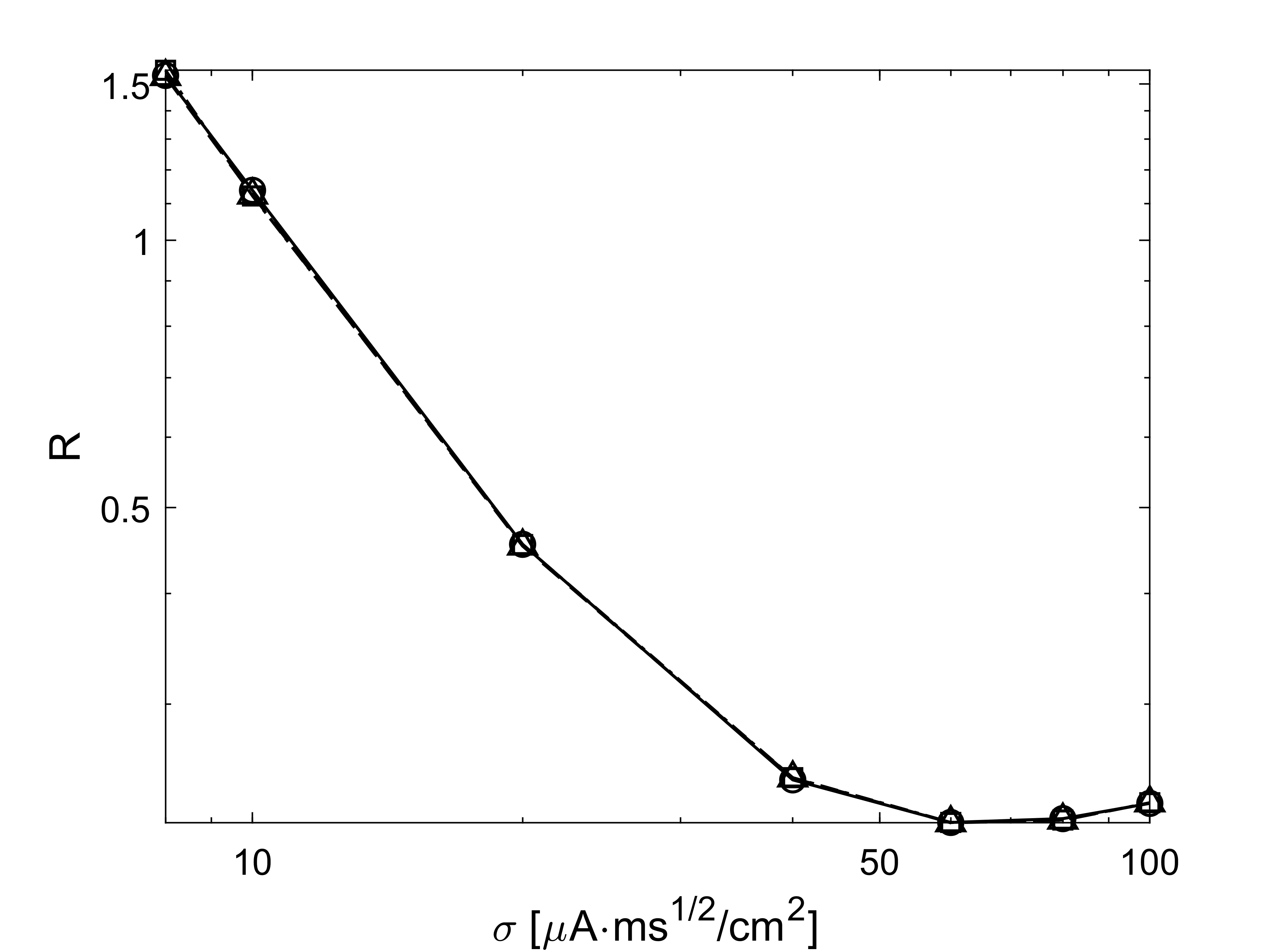}%
\caption{\label{fig:dCoS6p3} The dependence of $ R $ on $ \sigma $ for $ I = 6.2 $ $ \mu $A$/$cm$^{2}$. 
Different symbols (circles, triangles, and squares) represent the results for different sample paths of $ \xi (t) $.  
The lines (solid, broken, and dash-dot) are a guide for eyes. 
The irregularity $ R $ was estimated from a minimum of 15\,057 and a maximum of 71\,687 samples of the recurrence times.}
\end{figure}
Figure~\ref{fig:dCoS6p3} shows the dependence of the irregularity $ R $ on $ \sigma $ 
for different sample paths of $ \xi (t) $. 
The curves overlap well and are downward convex. 
The irregularity $ R $ is minimal at $ \sigma = 60 $ $ \mu $A$ \cdot $ ms$^{1/2}/$cm$^{2}$. 
This is exactly the coherence resonance. 

In studies of coherence resonance, 
the characteristic correlation time of the autocorrelation function 
is also used to evaluate the regularity~\cite{Pikovsky97,Luccioli06}. 
For a stationary stochastic process $ y(t) $, 
the characteristic correlation time $ \tau _{c} $ is defined as 
\begin{equation}
    \tau _{c} = \int _{0}^{\infty } C^{2} (t) dt .
\end{equation}
In this equation, 
$ C(t) $ is the autocorrelation function of $ y(t) $ and is given by 
\begin{equation}
    C(\tau ) = \frac{\left\langle (y(t) - \left\langle y \right\rangle )(y(t+\tau ) - \left\langle y \right\rangle ) \right\rangle}{\left\langle (y(t) - \left\langle y \right\rangle ) ^{2}  \right\rangle} ,   \label{eq:ace} 
\end{equation} 
where $ \tau $ represents the time difference. 
The characteristic correlation time $ \tau _{c} $ takes a larger value as $ y(t) $ shows a more regular oscillation. 

In the stationary state, 
for the membrane potential, 
the autocorrelation function corresponding to Eq.~(\ref{eq:ace}), $ C_{v} (\tau ) $, is given by 
\begin{equation}
    C_{v} (\tau ) = \frac{\left\langle ( v^{*}(\omega ) - \left\langle v^{*}(\omega ) \right\rangle )(v^{*}(\tau ,\omega ) - \left\langle v^{*}(\omega ) \right\rangle ) \right\rangle}{\left\langle (v^{*}(\omega ) - \left\langle v^{*}(\omega ) \right\rangle ) ^{2}  \right\rangle} .
\end{equation}
In this equation, 
$ v^{*}(\omega ) $ and $ v^{*}(\tau ,\omega ) $ are given by
\begin{eqnarray}
    v^{*}(\omega ) &=& \lim_{t \to \infty } \varphi _{hh} (t,\theta _{-t} \omega ) v_{0},  \\
    v^{*}(\tau ,\omega ) &=&  \varphi _{hh}(\tau ,\omega ) v^{*}(\omega ),  
\end{eqnarray}
where $ v_{0} $ is the initial condition of $ v $ and $ \varphi _{hh} $ is a map that provides the solution of 
Eqs.~(\ref{eqn:HH}) to (\ref{eqn:bnHHsqd}). 
On the other hand, 
for a given noise sample path $ \omega $, 
the time-averaged autocorrelation function of $ v(t,\omega ) $, $ \overline{C} _{v} (\tau , \omega ) $, can be defined as 
\begin{equation}
    \overline{C} _{v} (\tau , \omega ) = \lim_{T' \to \infty } \frac{1}{T'} \int _{0}^{T'} v^{*}(t, \omega ) v^{*}(t+\tau ,\omega ) dt. 
\end{equation}

\begin{figure}
\includegraphics[width=8.6cm]{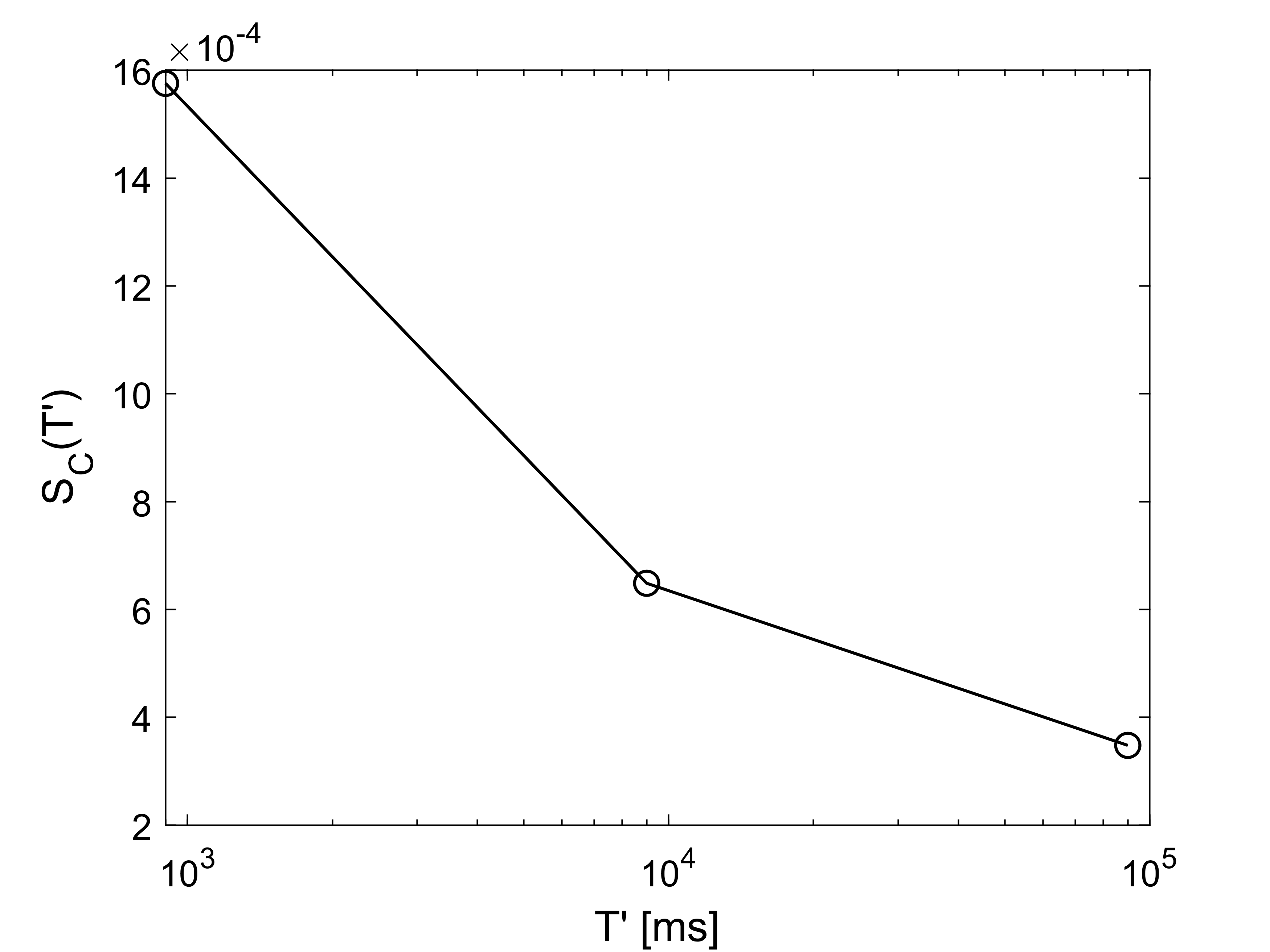}%
\caption{\label{fig:acer} The dependence of $ S_{C}(T') $  
on $ T' $. 
$ I = 6.2 $ $ \mu $A$/$cm$^{2}$ and $ \sigma = 10 $  $ \mu $A$ \cdot $ ms$^{1/2}/$cm$^{2}$. 
The solid line is a guide for eyes.}
\end{figure}
Figure~\ref{fig:acer} shows the dependence of the standard deviation  $ S_{C} (T') $ on $ T' $.  
Here, 
the standard deviation $ S_{C} (T') $ is given by 
\begin{equation}
    S_{C} (T') = \lim_{\tau _{u} \to \infty } \frac{1}{\tau _{u}} \int_{0}^{\tau_{u} } \left\langle \left( \widehat{ \overline{C}} _{v} (\tau ,T', \omega ) - C_{v} (\tau ) \right) ^{2} \right\rangle d\tau , 
\end{equation}
where $ \widehat{ \overline{C}} _{v} (\tau ,T', \omega ) $ is an estimate of $ \overline{C} _{v} (\tau , \omega ) $ and 
is given by 
\begin{equation}
    \widehat{ \overline{C}} _{v} (\tau ,T', \omega ) = \frac{1}{T'} \int _{0}^{T'} v^{*}(t, \omega ) v^{*}(t+\tau ,\omega ) dt. 
\end{equation} 
We can see that 
the standard deviation monotonically decreases as $ T' $ increases. 
This result suggests that $ \overline{C} _{v} (\tau , \omega ) $ is independent of $ \omega $ and 
is equal to $ C_{v} (\tau ) $. 

\begin{figure}
\includegraphics[width=8.6cm]{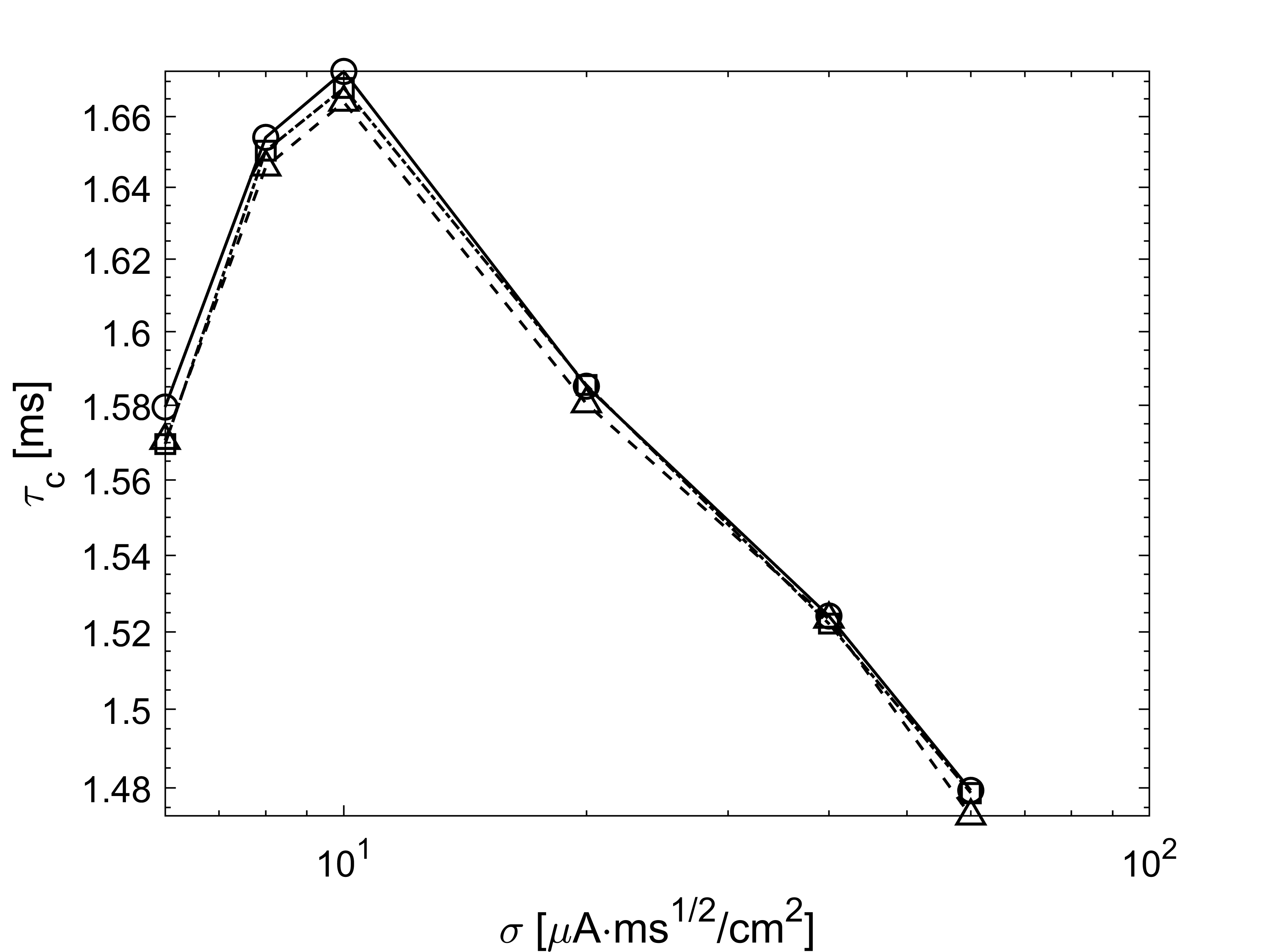}%
\caption{\label{fig:cds} The dependence of $ \tau _{c} $ on $ \sigma $ for $ I = 6.2 $ $ \mu $A$/$cm$^{2}$. 
Different symbols (circles, triangles, and squares) represent the results for different sample paths of $ \xi (t) $.  
The lines (solid, broken, and dash-dot) are a guide for eyes. 
The autocorrelation functions of $ v(t,\omega ) $ were estimated from simulations lasting a minimum of 1\,000\,000 ms and a maximum of 8\,000\,000 ms, 
from which the starting 2\,000 ms were discarded.}
\end{figure}
In the following, 
the characteristic correlation time for $ v(t) $  
is evaluated based on the time-averaged autocorrelation function of $ v(t,\omega ) $. 
Figure~\ref{fig:cds} shows the dependence of the characteristic correlation time $ \tau _{c} $ on $ \sigma $ 
for different sample paths of $ \xi (t) $.  
The curves overlap well and are upward convex. 
The characteristic correlation time $ \tau _{c} $ is maximal at $ \sigma = 10 $ $ \mu $A$ \cdot $ ms$^{1/2}/$cm$^{2}$. 
This is also the coherence resonance. 

\section{Discussion} \label{sec:discuss}
In the present study, 
we examined coherence resonance for time-averaged measures of the regularity.  
In the examination, 
we used a Hodgkin-Huxley neuron model  
driven by a constant current and a noise. 
We showed that for large times, 
the neuron is in a stationary state irrespective of initial conditions. 
We then showed numerical evidence that in the stationary state, 
a given noise sample path uniquely determines the dynamics of the neuron.  
We then showed numerical evidence suggesting that 
the time-averaged measures of the regularity of the dynamics 
is independent of noise sample paths and equal to the ensemble-averaged measures. 
In addition, 
we demonstrated coherence resonance for time-averaged measures. 

The irregularity of an oscillation $ R $  
is equivalent to 
the coefficient of variation (CV) of interspike intervals, which  
is often used in the study of coherence resonance~\cite{Pikovsky97,Luccioli06}. 
We denote the value of $ v $ on the Poincar{\'e} section as $ v_{p} $.  
We define the generation time of action potential as the time 
when the membrane potential $ v $ exceeds $ v_{th} $. 
When $ v_{th} = v_{p} $, 
the intervals between the generation times (interspike intervals) are equal to the recurrence times to the Poincar{\'e} section.  
In addition, 
the CV of interspike intervals is defined as the standard deviation of interspike intervals divided by 
the mean of interspike intervals. 
Thus, 
$ R $ is equal to CV of interspike intervals. 
This relation clarifies a nonlinear dynamical meaning of CV of interspike intervals: 
the variation of the recurrence times to a Poincar{\'e} section. 

A Poincar{\'e} section is usually used to construct a Poincar{\'e} map. 
The Poincar{\'e} map is useful for clarifying qualitative properties of 
$ (n+1) $-dimensional continuous dynamical systems governed by differential equations 
using $ n $-dimensional discrete dynamical system theory. 
On the other hand, 
in the present study, 
the Poincar{\'e} section is used to reduce the four-dimensional continuous 
stochastic process to a one-dimensional point process. 
This reduction enables us to characterize the dynamics of high-dimensional 
continuous dynamical systems from the properties of low-dimensional 
simple stochastic processes.    

In studies of coherence resonance, 
not only CV of interspike intervals and the characteristic correlation time but also 
the power spectrum is used to evaluate the regularity of the dynamics~\cite{Lee98}. 
In the present study, 
numerical evidence suggested that 
the time-averaged autocorrelation function is independent of noise sample paths and 
equal to the ensemble-averaged autocorrelation function.  
The same is true for the power spectrum because 
the autocorrelation function and the power spectrum have a one-to-one relation by 
the Wiener-Khinchin theorem.

\appendix
\section{Proof of Eq.~(\ref{eq:auinv})}    \label{app:inv} 
Here, 
we prove Eq.~(\ref{eq:auinv}).   
Substituting $ x_{0} = \sigma \int_{-\infty }^{0} e^{\gamma s} \xi (s;\omega ) ds $ into 
the right hand side of Eq.~(\ref{eq:irexamsol}) leads to 
\begin{eqnarray}
    \varphi (t,\omega ) x^{*} (\omega ) &=&  \sigma e^{-\gamma t} \int_{-\infty }^{0} e^{\gamma s} \xi (s;\omega ) ds \nonumber \\ 
& &+ \sigma \int _{0}^{t} e^{-\gamma (t-s)} \xi (s;\omega ) ds  \nonumber \\ &=& \sigma \int _{-\infty }^{t} e^{-\gamma (t-s)} \xi (s;\omega ) ds \nonumber \\
&=& \sigma \int _{-\infty }^{0} e^{\gamma s} \xi (s+t;\omega ) ds \nonumber \\
&=& x^{*} (\theta _{t} \omega ).  \nonumber 
\end{eqnarray}

\section{Boundedness of the pullback of $ v $}    \label{app:bv}
The boundedness of the pullback of $ v $ has already been proven for $ I = 0 $ in Ref.~\cite{Pakdaman01}.  
The boundedness of the pullback of $ v $ for $ I > 0 $ can also be proven in the same way. 
Eq.~(\ref{eqn:HH}) with $ C = 1 $ $\mu $F$/$cm$^{2}$ can be rewritten as 
\begin{equation}
    \frac{dv}{dt} = G(t)(a(t)-v) + I + \sigma \xi (t), 
\end{equation}
where $ G(t) $ and $ a(t) $ are given by 
\begin{eqnarray}
    G(t) &=& g_{Na}m^{3} h + g_{K}n^{4} + g_{L},    \\ 
    a(t) &=& \frac{g_{Na}m^{3} h V_{Na} + g_{K}n^{4} V_{K} + g_{L}V_{L}}{G(t)}.  
\end{eqnarray}
Here, 
we define $ v' $ as $ v'(t) = \int_{-\infty }^{t} \exp (-\int_{s}^{t} G(r) dr) G(s) a(s) ds $. 
This is a solution of the equation: 
\begin{equation}
    \frac{dv'}{dt} = G(t)(a(t)-v'). 
\end{equation}
We also define $ u $ as 
\begin{equation}
    u(t,\omega ) = \int_{-\infty }^{t} e^{-\gamma (t-s)} \left( I + \sigma \xi (s,\omega ) \right) ds. \label{apeq:u}
\end{equation}
This is a solution of the equation: 
\begin{equation}
    \frac{du}{dt} = -\gamma u + I + \sigma \xi (t).    \label{eq:ou} 
\end{equation}
When we define $ y $ as $ y = v-v'-u $,  
we have 
\begin{equation}
    \frac{dy}{dt} = -G(t) y + (\gamma -G(t))u(t,\omega ).   
\end{equation}
From this equation, 
we have 
\begin{equation}
    \frac{dy^{2}}{dt} = -2G(t) y^{2} + 2(\gamma -G(t))u(t,\omega )y.    \label{eq:appy2}   
\end{equation}
Thus, 
we have 
\begin{equation}
    \frac{dy^{2}}{dt} \le -\beta y^{2} + \zeta \left[ u(t,\omega ) \right] ^{2},    \label{eq:ine}
\end{equation}
where $ \beta $ and $ \zeta $ are positive constants.  
From Eq.~(\ref{eq:ine}) and the comparison theorem, 
we have 
\begin{equation}
    \phi (t,\omega ) y_{0} \le e^{-\beta t} y_{0}^{2} + \zeta \int_{0}^{t} e^{-\beta (t-s)} \left[ u(s,\omega ) \right] ^{2} ds,  \label{eq:appb1}
\end{equation}
where $ \phi (t,\omega ) $ is the map from the initial condition $ y_{0} $ to $ y^{2}(t,\omega ) $. 
The map is defined by the solution of Eq.~(\ref{eq:appy2}).  
From Eq.~(\ref{eq:appb1}), 
we have 
\begin{equation}
    \lim_{t \to \infty } \phi (t,\theta _{-t} \omega ) y_{0} \le \zeta \int_{-\infty }^{0} e^{\beta s} \left[ u(s,\omega ) \right] ^{2} ds. \label{eq:inv2}
\end{equation}
Eq.~(\ref{eq:inv2}) shows that there is a bound 
for the pullback of $ y $. 
The pullback of $ v' $ is also bounded because 
$ 0 < g_{L} < G(t) < g_{Na} + g_{K} + g_{L} $ and min$(V_{rev}) < a(t) < $max$ (V_{rev}) $, 
where $ V_{rev} = V_{Na} $, $ V_{K} $, or $ V_{L} $. 
From Eq.~(\ref{apeq:u}),  
the pullback of $ u $ is given by $ I/\gamma + \sigma \int_{-\infty }^{0} e^{\gamma s} \xi (s) ds $.  
Thus, 
the pullback of $ u $ is bounded.  
Given that $ v=y+v'+u $, 
the pullback of $ v $ is bounded.

\begin{acknowledgments}
This work was supported by Tokyo Metropolitan Government Advanced Research Grant R2-2. 
\end{acknowledgments}


\begin{thebibliography}{99}
\bibitem{Gang93} H. Gang, T. Ditzinger, C. Z. Ning, and H. Haken, Phys. Rev. Lett. {\bf 71}, 807 (1993).
\bibitem{Longtin97} A. Longtin, Phys. Rev. E {\bf 55}, 868 (1997). 
\bibitem{Pikovsky97} A. S. Pikovsky and J. Kurths, Phys. Rev. Lett. {\bf 78}, 775 (1997). 
\bibitem{Lee98} S. G. Lee, A. Neiman, and S. Kim, Phys. Rev. E {\bf 57}, 3292 (1998). 
\bibitem{Dubbeldam99} J. L. A. Dubbeldam, B. Krauskopf, and D. Lenstra, Phys. Rev. E {\bf 60}, 6580 (1999). 
\bibitem{Buldu01} J. M. Buld\'{u}, J. Garc\'{i}a-Ojalvo, C. R. Mirasso, M. C. Torrent, and J. M. Sancho, Phys. Rev. E {\bf 64}, 051109 (2001). 
\bibitem{Hizanidis08} J. Hizanidis and E. Sch\"{o}ll, Phys. Rev. E {\bf 78}, 066205 (2008).
\bibitem{Ushakov05} O. V. Ushakov, H.-J. W\"{u}nsche, F. Henneberger, I. A. Khovanov, L. Schimansky-Geier, and M. A. Zaks, Phys. Rev. Lett. {\bf 95}, 123903 (2005). 
\bibitem{Arteaga07} M. Arizaleta Arteaga, M. Valencia, M. Sciamanna, H. Thienpont, M. L\'{o}pez-Amo, and K. Panajotov, Phys. Rev. Lett. {\bf 99}, 023903 (2007). 
\bibitem{Kabiraj15} L. Kabiraj, R. Steinert, A. Saurabh, and C. Oliver Paschereit, Phys. Rev. E {\bf 92}, 042909 (2015). 
\bibitem{Mompo18} E. Mompo, M. Ruiz-Garcia, M. Carretero, H. T. Grahn, Y. Zhang, and L. L. Bonilla, Phys. Rev. Lett. {\bf 121}, 086805 (2018).   
\bibitem{Zhu19} Y. Zhu, V. Gupta, and L. K. B. Li, J. Fulid Mech. {\bf 881}, R1 (2019). 
\bibitem{Luccioli06} S. Luccioli, T. Kreuz, and A. Torcini, Phys. Rev. E {\bf 73}, 041902 (2006). 
\bibitem{Abbott90} L. F. Abbott and T. B. Kepler, in {\it Statistical Mechanics of Neural Networks}, {\it Barcelona}, {\it 1990}, edited by L. Garrido (Springer, Berlin, 1990), pp.5-18. 
\bibitem{Arnold98} L. Arnold, {\it Random Dynamical Systems}, (Springer-Verlag, Berlin, Heidelberg, 1998). 
\bibitem{Crauel99} H. Crauel and M. Gundlach, {\it Stochastic Dynamics}, (Springer, New York, 1999).
\bibitem{Caraballo17} T. Caraballo and X. Han, {\it Applied Nonautonomous and Random Dynamical Systems}, (Springer, Cham, 2017). 
\bibitem{Pakdaman01} K. Pakdaman and S. Tanabe, Phys. Rev. E {\bf 64}, 050902 (2001). 
\end{thebibliography}

\end{document}